\documentclass[showpacs,preprintnumbers,amsmath,amssymb,twocolumn,nobibnotes,longbibliography,nofootinbib,superscriptaddress]{revtex4-2}
\usepackage[english]{babel}
\usepackage{mathtools}
\usepackage{graphicx}
\usepackage{mathrsfs}
\usepackage{amssymb}
\usepackage{xcolor}
\usepackage[colorlinks=true,citecolor=blue,urlcolor=blue]{hyperref}
\usepackage{amsmath}
\usepackage{bm}
\usepackage{physics}

\usepackage{ulem}

\begin{document}

\title{ Chirality Driven Ratchet Currents in Two-Dimensional Tellurene with an Asymmetric Grating	}

\author{M. D. Moldavskaya}
\affiliation{Physics Department, University of Regensburg, 93040 Regensburg, Germany}
\author{L. E. Golub}
\affiliation{Institute of Theoretical Physics and Halle-Berlin-Regensburg Cluster of Excellence CCE, University of Regensburg, 93040 Regensburg, Germany}
\author{Chang Niu}
\affiliation{Elmore Family School of Electrical and	Computer Engineering, Purdue University, West Lafayette, Indiana 47907, United States}
\affiliation{Birck Nanotechnology Center, Purdue University, West Lafayette, Indiana 47907, United
	States}
\author{Peide D. Ye}
\affiliation{Elmore Family School of Electrical and	Computer Engineering, Purdue University, West Lafayette, Indiana 47907, United States}
\affiliation{Birck Nanotechnology Center, Purdue University, West Lafayette, Indiana 47907, United
	States}
\author{S. D. Ganichev}
\affiliation{Physics Department, University of Regensburg, 93040 Regensburg, Germany}
\affiliation{CENTERA Labs, Institute of High Pressure Physics of the Polish Academy of Sciences, 01-142 Warsaw, Poland}
\email{sergey.ganichev@ur.de}


		\begin{abstract}			
The emergence of the terahertz (THz) ratchet effect is a rapidly expanding field of research that utilizes broken spatial symmetry in low-dimensional materials to rectify alternating current (AC) induced by THz fields into direct current (DC). This mechanism is highly promising for next-generation, room-temperature terahertz applications, particularly in high-speed, sensitive detection and imaging.  In this work, we explore a ratchet effect generated in two dimensional tellurene, a novel promising  semiconductor material  consisting of helical atomic chains, creating a structure with inherent chirality.  As a key result, the DC circular ratchet current flowing in the chiral axis direction $c$ is determined by the helicity of the radiation and can be reversed by switching the helicity from right to left handed. The circular ratchet effect excited by THz laser radiation is demonstrated for room temperature.  The effect is demonstrated at various gate voltages when the Fermi level lies in vicinity of the Weyl point in the conduction band, in the band gap, and in the valence band with almost  parabolic energy dispersion. The results are described by the developed microscopic theory  based on the Boltzmann kinetic equation approach. 

		\end{abstract}
	
				\maketitle

	
Chirality is a profound and ubiquitous concept that permeates various realms of nature, manifesting itself in an extensive array of materials and phenomena. From the intricate molecular biology that underpins all living organisms to the fundamental principles of particle physics, chirality plays a pivotal role in shaping our understanding of the natural world. At its core, chirality refers to the unique spatial asymmetry found in certain structures that cannot be superimposed by any combination of rotation and translation. The significance of chirality goes beyond mere structural differences. There are also chirality-dependent chemical and physical properties, as well as different responses to external stimuli, see Refs.~\cite{Crassous2023,Zhu2024,Matassa2024,Habibovic2024} for reviews. In condensed matter physics, it is bulk tellurium and two-dimensional (2D) tellurene which are most well studied chiral materials~\cite{Calavalle2022,Moldavskaya2023,Slawinska2023,Joseph2024,SuarezRodriguez2024}. Important example for opto-electronic phenomena in bulk Te are the circular photogalvanic and circular photon drag effects, where an electric current is driven by the light helicity~\cite{Asnin1978,Shalygin2016,Moldavskaya2023}. Most recently, circular photogalvanic photocurrents have been also detected in two-dimensional (2D) tellurene~\cite{Niu2023,Moench2024}. 
The emergence of 2D Te-based structures provides a new opportunity to study chirality-dependent optoelectronic phenomena, in which {\it extrinsic} chirality effects are produced by superimposing an asymmetrical grating on 2D tellurene samples, which creates ratchet structures. This unifies the physics of chiral materials with the ratchet effect --- the generation of a DC electric current in response to an AC electric field in systems with broken inversion symmetry, for reviews see Refs.~\cite{Linke2002,Reimann2002,Haenggi2009,Ivchenko2011,Denisov2014,Bercioux2015,Cubero2016, Reichhardt2017}. In this work, we propose and demonstrate that a THz electric field acting on an asymmetrical grating fabricated on a 2D tellurene sample can efficiently generate a ratchet current. The direction of this current reverses when the light helicity changes from left- to right-circular polarization. We also developed a microscopic theory describing the formation of this current for Weyl and parabolic   energy bands, which are relevant for positive and negative back gate voltages applied to the tellurene layer.

\begin{figure*}[t]
	\centering
	\includegraphics[width=\linewidth]{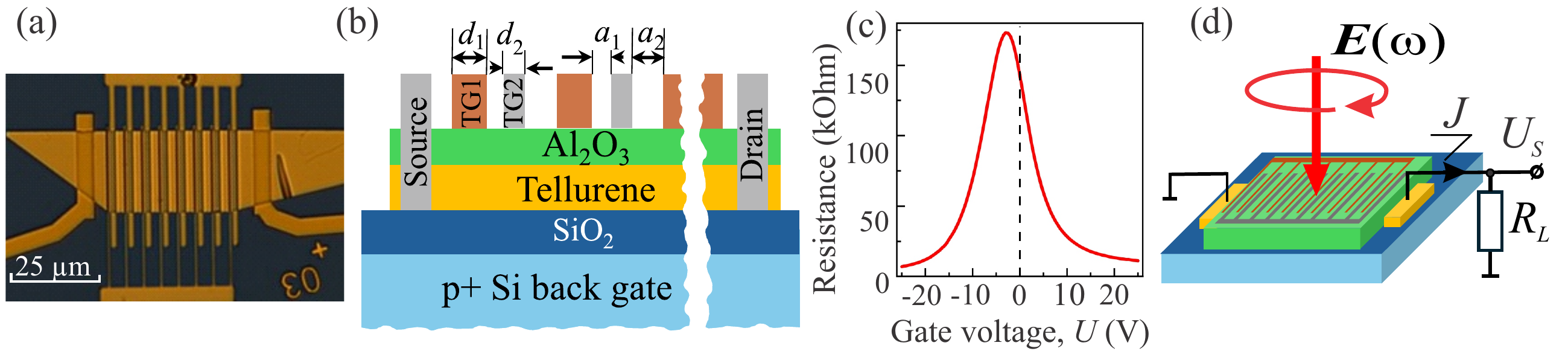}
	\caption{Panels (a) and (b) show  an optical micrograph and a cross-section of the ratchet structure. Tellurene flake is deposited on top of $p+$ Si substrate covered by 90~nm SiO$_2$ layer. Nickel  contacts  were fabricated by electron beam evaporation on the short sides of the rectangular flakes, which were $46 \times 23~\mu$m$^2$. 
		To cap the Te flakes, we deposited a 15-nm-thick layer of Al$_2$O$_3$. 
		The asymmetric grating consists of supercells, each formed by two Ti stripes having widths $d_1 = 1.5$ and  $d_2 = 0.5$~$\mu$m,   spacings  $a_1 = 0.5$ and $a_2=2.5$~$\mu$m, and thickness~50~nm. 
		The supercell is  repeated $8$ times forming the asymmetric grating  with the period $d = d_1 + a_1 +d_2 +a_2$. Panel (c) shows the two-point resistance as a function of the back gate voltage. Panel (d) sketches the experimental set-up. 
	}
	\label{fig1}
\end{figure*}

\section{Basic concept}

The proposed ratchet structure under study consists of a 2D tellurene conduction layer with a superimposed asymmetric lateral metallic grating, as shown in Fig.~\ref{fig1}. The removal of spatial inversion in the structure is a necessary condition for the ratchet effect, and the asymmetrical lateral superlattice serves as an extrinsic source for this effect. The grating results in the electrostatic potential $V(x,y)$, acting on the electrons being periodic in the $x$ direction with a period $d$: $V(x+d,y)=V(x,y)$. In equilibrium, the electrons occupy the potential minima, resulting in modulated electron density with a period $d$. When uniform THz radiation is normally incident on an asymmetric grating, an additional periodically modulated electric field is produced that acts on the electrons in the tellurene layer. This field is caused by the near field of diffraction and is enhanced in comparison with the far field of radiation. Consequently, the amplitude of the electric field, $\bm E(\bm r,t)=\bm e E(\bm r)\exp(-i\omega t)+c.c.$, is modulated in space with the same period: $E(x+d,y)=E(x,y)$. Here, $\bm e$ is the polarization vector, which can be complex for elliptical polarization, and $\omega$ is the radiation frequency. The modulation results in the simultaneous action of both the static and dynamic forces:
\begin{equation}
\label{F}
\bm F(\bm r, t) = - \bm \nabla V(\bm r) + e \bm E(\bm r,t).
\end{equation}
which causes the ratchet DC electric current. 
For the modulation period $d$ much larger than the electron mean free path and for the photon energy and the potential size much smaller than the electron Fermi energy, we can use the kinetic theory for calculations of the ratchet currents. The Boltzmann equation for the distribution function $f(\bm p, \bm r, t)$ has the form
\begin{equation}
	\label{kin_eq}
	\pdv{f}{t} + \bm v_{\bm p} \cdot \pdv{f}{\bm r} + \bm F(\bm r, t) \cdot \pdv{f}{\bm p} = {\rm St}[f].
\end{equation}
Here $\bm v_{\bm p} = \partial \varepsilon_{\bm p}/\partial \bm p$ is the group velocity of electrons with the momentum $\bm p$ and energy $\varepsilon_{\bm p}$, $\rm St$ is the collision integral, and the force acting on the electrons is a sum of the static and time-oscillating terms, Eq.~\eqref{F}. In order to get the ratchet current we iterate the kinetic Eq.~\eqref{kin_eq} in small perturbations $V$ and $E$ and find the DC correction to the distribution function $\delta f(\bm p, \bm r) \propto E^2\nabla_{x,y}V$. Then the current density is calculated as
\begin{equation}
	\bm j = eg_s g_v\sum_{\bm p} \bm v_{\bm p} \, \overline{\delta f(\bm p, \bm r)},
\end{equation}
where the factors $g_s$ and $g_v=2$ account for the spin and valley degeneracy in tellurene ($g_s=1$ for the valence band and $g_s=2$ for the conduction band), and the line denotes averaging over coordinates. Consequently, the ratchet current is proportional to the components $\Xi_{x,y}$ of  the lateral asymmetry parameter
\begin{equation}
\bm \Xi = \overline{E^2(\bm r)\bm \nabla V(\bm r)}.
\end{equation}

Despite both the static force $-\bm \nabla V$ and the radiation intensity $I$ modulation $\propto E^2$ have zero mean values, their averaged product $\bm \Xi$ is nonzero in the studied structure. Figure~\ref{fig1}(a) shows that the  structure has the lowest possible symmetry, $C_1$, without any nontrivial symmetry elements. In particular, the structure is chiral because it has no mirror reflection planes, and both components $\Xi_x$ and $\Xi_y$ are nonzero. The symmetry analysis shows that the DC electric current can be driven by circularly-, linearly-polarized or unpolarized radiation
\begin{equation}
	\label{jx}
j_x= \Xi_y \gamma P_{\rm circ} +\Xi_x \left( \chi_L  {P_L} + \chi_0 \right) +\Xi_y \tilde{\chi}_L {\tilde{P}_L}.
\end{equation}
Here $P_{\rm circ}$ is the radiation helicity, and $P_L$, $\tilde{P}_L$ are the Stokes parameters defining the degrees of the linear polarization~\cite{Saleh2019}, and the factors $\gamma$, $\chi_L$, $\tilde{\chi}_L$, $\chi_0$ describe the circular, linear and polarization-independent  ratchet effects, respectively. These factors are independent of the periodic potential and radiation intensity, being determined by the properties of 2D carriers in tellurene and radiation frequency. As an important result, the first term in Eq.~\eqref{jx} shows that the helicity-driven ratchet current in tellurene-based devices is expected in the direction $x$ corresponding to the chiral axis $c$.

\section{Devices and experimental technique}

To demonstrate that the circular ratchet effect emerges in tellurene-based devices, we fabricated an asymmetric metal interdigitated grating on a tellurene flake covered by Al$_2$O$_3$ for electrical isolation.  Figures~\ref{fig1}(a) and~(b) show the photograph and cross-section of the studied devices, respectively. The 2D Te flakes were synthesized using the hydrothermal growth method~\cite{Qiu2020}.
The synthesized 2D Te flakes, which had a thickness of approximately 30~nm, were rinsed twice in water before being transferred to a 90-nm SiO$_2$/Si substrate using the Langmuir-Blodgett method. This substrate served as a back gate. The long side of the tellurene bar is oriented along the $c$-axis of the tellurium.
		In the following, the long and short sides correspond to the $x$- and $y$-directions, respectively. 
To cap the Te flakes, we deposited a 15-nm-thick layer of Al$_2$O$_3$ on top using atomic layer deposition with (CH$_3$)$_3$Al (TMA) and H$_2$O as precursors at 200$^\circ$C.

Asymmetric grating was formed by two combs having different stripe widths and separations, see~Fig~\ref{fig1}. To demonstrate that the circular photocurrent is caused by exclusively the ratchet effect, we additionally fabricated tellurene flakes with the same technology, but without grating. The experiments described below were performed at room temperature.  To change the concentration and carrier type, a voltage $U$ is applied to the back gate. Sweeping the back gate voltage shows that the charge neutrality point (CNP) is observable in the two-point resistance, as seen in Fig.~\ref{fig1}(c). For the investigated devices, the CNP was found at negative gate voltages of approximately $-3$~V. For subsequent analysis, the back gate voltage is presented as an effective voltage, $U_{\rm G} = U - U_{\rm CNP}$, where $U_{\rm CNP}$ is the voltage at which the CNP appears. Electron and hole-type conductivity occur for positive and negative $U_{\rm G}$, respectively, as described in Ref.~\cite{Niu2023a}. The carrier concentration for an effective gate voltage of $\pm 20$~V is about $5 \times 10^{12} $~cm$^{-2}$, and the mobility $\mu$ is about $520$ and 200~cm$^2$/Vs for holes and  electrons, respectively.

Tellurene-based devices were excited by normal-incident THz laser radiation, see Fig.~\ref{fig1}(d). As a source of radiation we used optically pumped pulsed NH$_3$ laser operated at $f = 1.07$~THz ($\lambda = 280~\mu$m, $\hbar\omega =4.4$~meV) and 2.02~THz ($\lambda = 148~\mu$m, $\hbar\omega =8.35$~meV). The laser generated single pulses with a duration of about 100\,ns and a repetition rate of 1~Hz. The radiation with  $f \approx 1.07 (2.02)$~THz and power $P\approx 10 (30)$~kW was focused in a spot of about  5(3.8)~mm diameter. Taking into account the area of the device, $S_{\rm device} \approx 10^{-5}$~cm$^{-2}$,   we obtain that the power irradiating the sample $P_s = P\times (S_{\rm device}/S_{\rm spot})\approx 0.5 (2.7)$~W, where $S_{\rm spot}$ is the  beam spot area. The induced photocurrents were detected as a voltage drop  across load resistors $R_L=50$~Ohm.  The signals were recorded  using digital oscilloscopes.

\section{THz-induced circular ratchet currents}

\begin{figure}[t] 
	\centering
	\includegraphics[width=\linewidth]{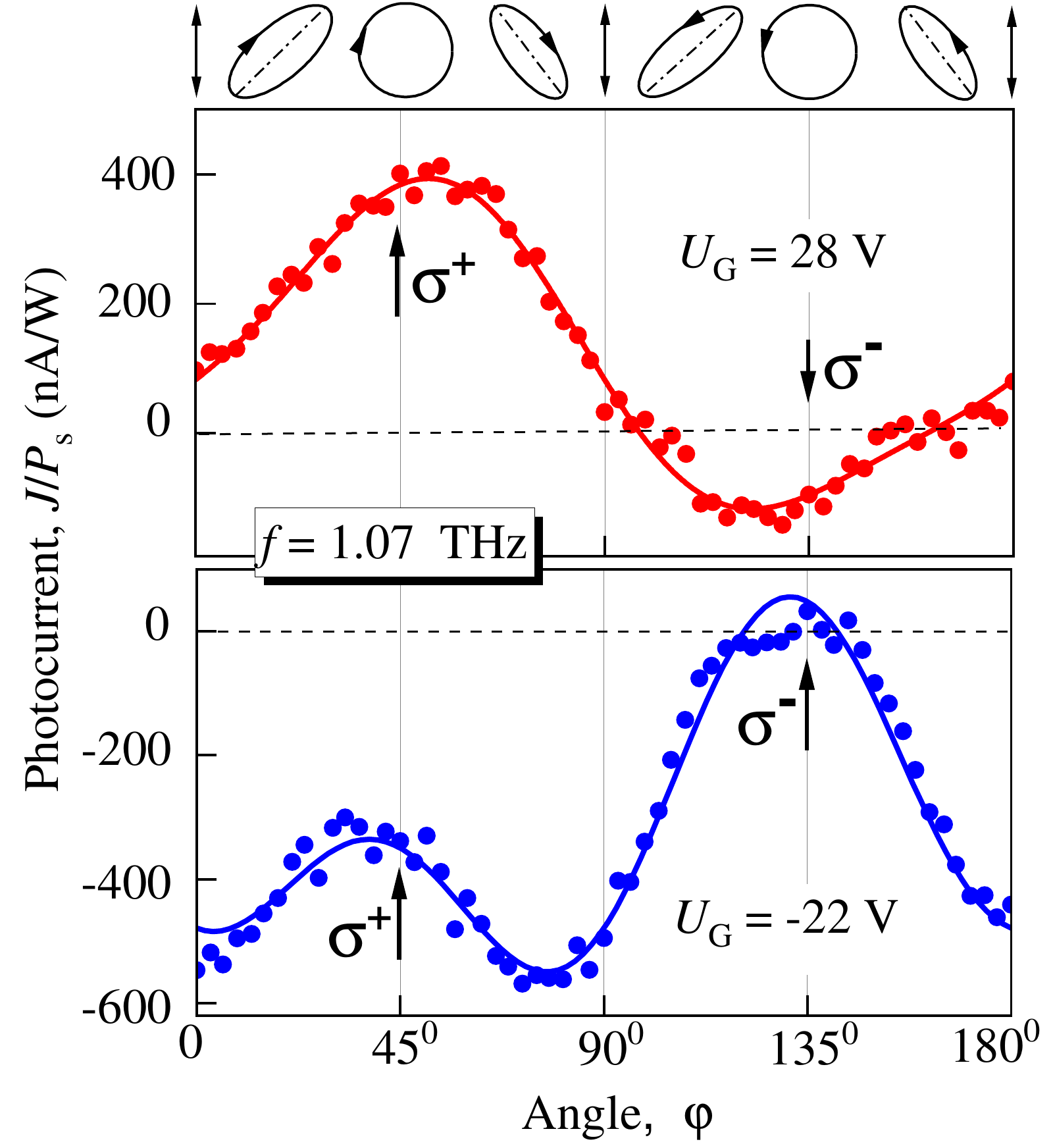}
	\caption{The normalized photocurrent, $J/P_{\rm s}$, as a function of the lambda-quarter rotation angle, $\varphi$. The data are shown for a  radiation frequency $f=1.07$~THz and several effective back gate voltages ranging from hole- ($U_{\rm G} < 0$) to electron-conductivity ($U_{\rm G} > 0$). The solid lines represent the corresponding fits according to Eq.~\eqref{phi}. The fitting coefficients are plotted in Fig.~\ref{fig4} and Fig.~S4 of SI.  Vertical arrows indicate the right- ($\sigma^+$) and left-handed ($\sigma^-$) circular polarizations, and the ellipses on top illustrate the polarization states at various angles~$\varphi$. 
	}
	\label{fig2}
\end{figure}

\begin{figure}[t] 
	\centering
	\includegraphics[width=\linewidth]{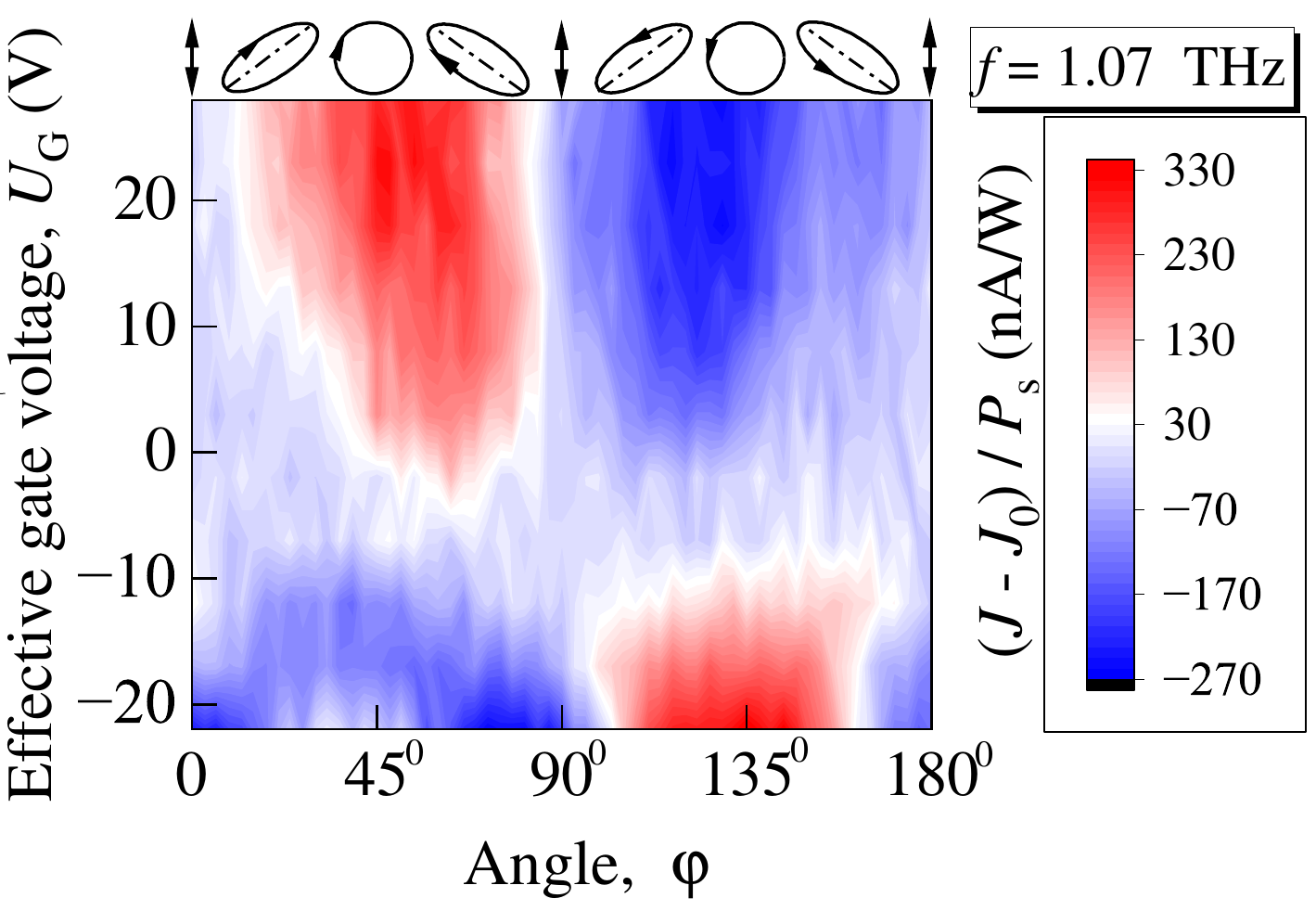}
	\caption{Phase angle $\varphi$ and effective gate voltage dependencies of the color-coded photocurrent strength $(J-J_0)/P_{\rm s}$  as a function of the angle $\varphi$. The data are obtained for $f=1.07$~THz. 
	}
	\label{fig3}
\end{figure}

\begin{figure}[h] 
	\centering
	\includegraphics[width=\linewidth]{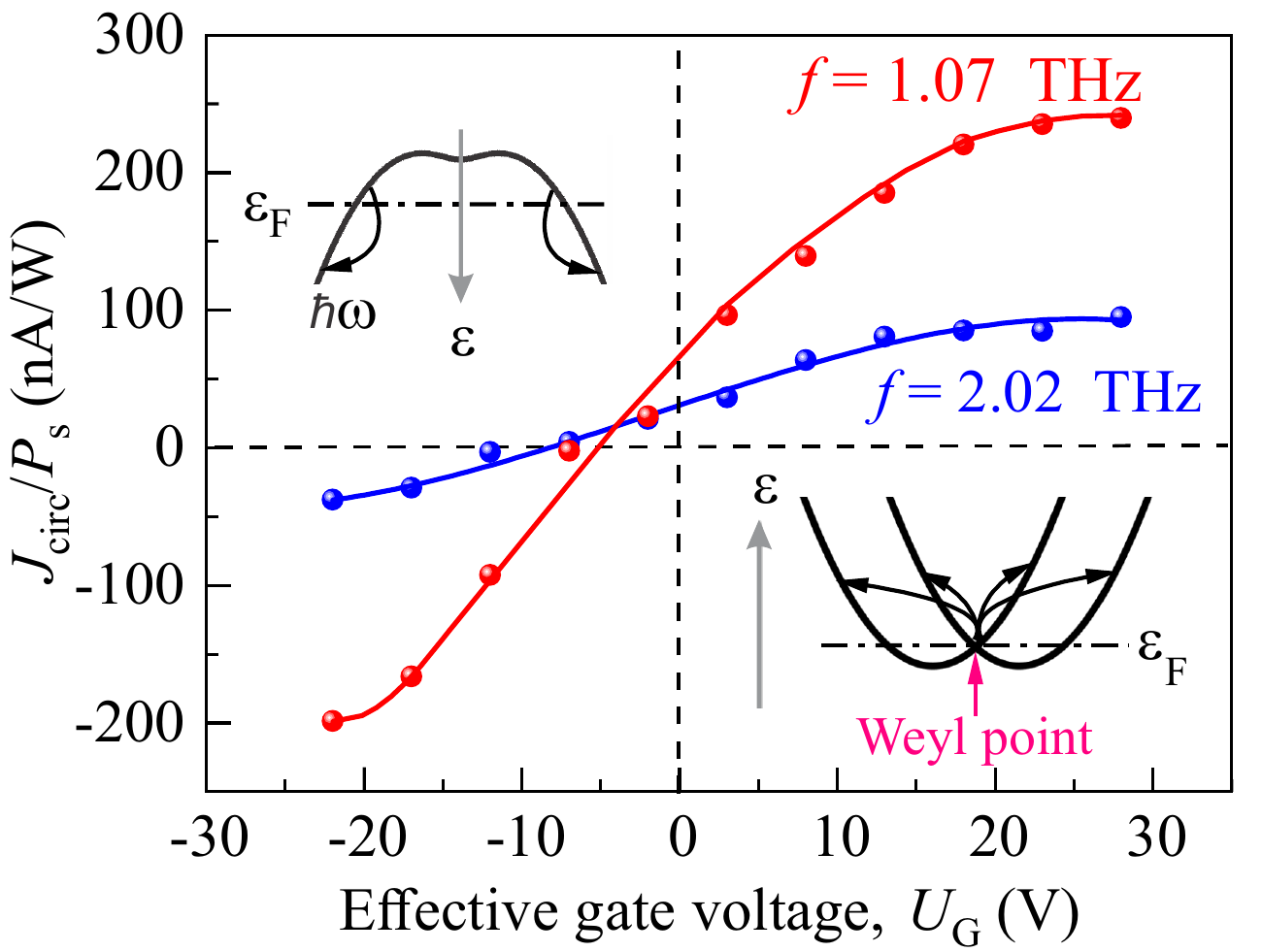}
	\caption{Dependencies of the circular photocurrent excited by radiation with frequencies $f=1.07$ and $f=2.02$~THz on the effective gate voltage  $U_{\rm G}$. Insets sketches indirect (Drude-like) optical transitions in the valence (left, negative gate voltages) and conduction (right, positive gate voltages) energy bands.
	}
	\label{fig4}
\end{figure}

\begin{figure}[h] 
	\centering
	\includegraphics[width=\linewidth]{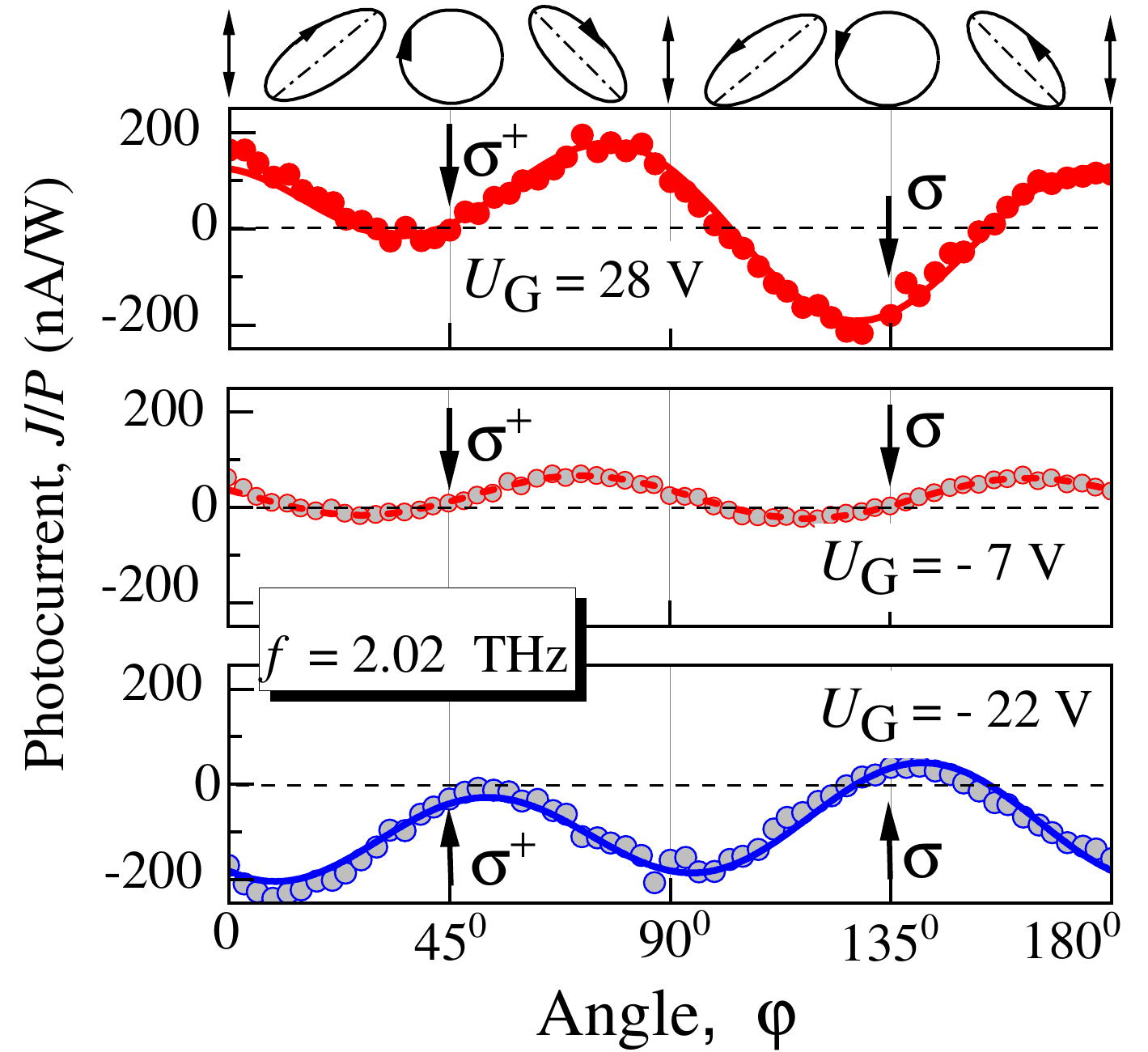}
	\caption{Dependencies of the photocurrent (colored symbols) excited by radiation with $f=2.02$~THz on the rotation angle $\varphi$ of the $\lambda/4$ wave plate. The data are shown for several effective back gate voltages ranging from hole ($U_{\rm G} < 0$) to electron conductivity ($U_{\rm G} > 0$).  The photoresponse is normalized to the laser power $P_{\rm s}$  incident on the sample.  The vertical arrows in each panel mark the right-handed ($\sigma^+$) and left-handed ($\sigma^-$) circular polarization states, respectively. The solid lines represent the corresponding fits according to Eq.~\eqref{phi}. The fit coefficients are plotted in Fig.~\ref{fig4} and Fig.~S5 of SI.
	}
	\label{fig5}
\end{figure}
We used a $\lambda/4$ plate to excite ratchet currents sensitive to radiation helicity. By rotating the plate, we varied the THz radiation helicity according to $P_{\rm circ} \propto\sin{2\varphi}$~\cite{Saleh2019,Belkov2005}, where $\varphi=0$ is the angle between the laser polarization plane and the optical axis of the plate.~\footnote{Note that for $\varphi=0$, the radiation is linearly polarized, with the electric field vector $\bm E$ tilted from the $y$-direction by about $15^\circ$. } Figures~\ref{fig2}  and~\ref{fig3} show the data obtained for different effective gate voltages when applying radiation at a frequency of 1.07~THz. 

As a central result, Figs.~\ref{fig2} and~\ref{fig3} reveal that the photocurrents excited by right ($\sigma^+$) and left ($\sigma^-$) circularly-polarized radiation have substantially different amplitudes.  We find that the total photocurrent is well fitted by above derived Eq.~\eqref{jx}, which for the used setup takes the form
\begin{align}
	J = J_{\rm circ}\sin(2\varphi) + J_0  + J_{\rm L} \cos4(\varphi +\varphi_0)\,
	\label{phi}
\end{align}
where $J_{\rm circ} \propto \Xi_y \gamma$ is the amplitude of the circular photocurrent, $J_0$ is the amplitude of the polarization-independent contribution, and $J_{\rm L}$ is the amplitude of the photocurrent in response to linearly polarized radiation.\,\footnote{In this equation we have merged both linear ratchet currents using that, in the current setup, the Stokes parameters take form $P_L=(1+\cos{4\varphi})/2$ and $\tilde{P}_L=\sin{4\varphi}$~\cite{Belkov2005}.} Figures~\ref{fig2} and~\ref{fig3} reveal the dominance of the circular photocurrent over the linear one. Importantly, in the reference sample without grating, the circular photocurrent is absent, see Fig.~S1 in the Supporting Information (SI).

The dependence of the circular ratchet current on the effective gate voltage is shown in Fig.~\ref{fig4}. It reveals that the current reverses its direction at small negative effective gate voltages. By varying the radiation intensity, we observed that the current changes according to $J\propto E^2$, as shown in Fig.~S2 in SI. Similar behavior is detected for radiation with a frequency about twice as high, $f=2.02$ THz, but with an essential contribution of the linear current $J_L$ over the entire range of gate voltages, see Figs.~\ref{fig4} and~\ref{fig5}, as well as Figs.~S3 and S5 in SI.

\section{Discussion and microscopic theory}

The observed circular ratchet current in tellurene-based devices, being in focus of our work, can be illustrated by the following model. The radiation electric field changes the spatially-modulated distribution of electrons resulting in the new, nonequilibrium modulation profile which is formed after a quarter of the period $\pi/(2\omega)$. At this moment, the electric field vector of the circularly polarized radiation is already rotated by $90^\circ$ from the initial direction, accelerating the electrons. After a half of the period, the situation tends to reverse, however, the electrons are bunched in the other parts of the spatially modulated structure, where the electric field amplitude has another value. As a result, the compensation does not take place giving rise to the DC ratchet current. Remarkably, at opposite circular polarization, the acceleration occurs in the opposite direction, and the circular ratchet current $J_{\rm circ} \propto P_{\rm circ}$ reverses. This is the mechanism of the circular ratchet effect.

Following to the theory developed above, the ratchet current scales linearly with the radiation intensity ${I \propto E^2}$ as observed in experiments, see insets in Fig.~S2 in SI. The  gate voltage dependence of the circular ratchet current is presented in Fig.~\ref{fig4}. Depending on the gate voltage  $U_{\rm G}$, the Fermi level lies in the conduction band, valence band or in the band gap, see insets in Fig.~\ref{fig4}. The conduction band which is relevant for transport at large positive  $U_{\rm G}$, consists of two Rashba-splitted subbands in each tellurene valley. At the bottom of conduction band, a Weyl point in the 2D bandstructure is present. Consequently, we deal with Weyl fermions at a range of positive gate voltages, and with an almost parabolic energy dispersion at negative voltages.

To understand the observed circular ratchet current behavior upon variation of the gate voltage and radiation frequency, we develop the microscopic theory of the effect. We show that the circular ratchet current 
\begin{equation}
	\label{j_gamma}
j_x^{\rm circ}=\gamma \Xi_y P_{\rm circ}
\end{equation}
 is strongly sensitive to the type of both 2D energy dispersion of charge carriers and the disorder scattering potential. 
First we consider  parabolic energy dispersion $\varepsilon_{\bm p}=p^2/(2m)$.
At degenerate statistics for the short-range (SR)  disorder scattering potential we obtain (see SI for derivation)
%
\begin{equation}
	\label{chi_C_SR}
	\gamma^{\rm SR} 
	= {e^3 \tau^2\over 2\pi\hbar^2 m\omega [1+(\omega-\omega_{\rm pl}^2/\omega)^2\tau^2]}.
\end{equation}
Here $\tau$ is the transport scattering time,
and $\omega_{\rm pl}=sq$ is the 2D plasmon frequency with $q=2\pi/d$ and $s$
being the plasmon wavevector and velocity, respectively. 
For the short-range disorder, $\tau$ is independent of the Fermi energy $\varepsilon_{\rm F}$.
For long-range Coulomb (Coul) disorder scattering
potential we get
\begin{equation}
	\label{chi_C_Coul}
	\gamma^{\rm Coul} = \gamma^{\rm SR}{3\Omega^6 + 29\Omega^4 + 4\Omega^2 +32\over (1+\Omega^2)(\Omega^2+4)^2},
\end{equation}
where $\Omega=\omega\tau$, and $\tau \propto \varepsilon_{\rm F}$.

For the linear  energy dispersion $\varepsilon_{\bm p}=v_0p$, relevant to the vicinity of the Weyl point (at small positive  $U_{\rm G}$), the expressions for $\gamma^{\rm SR}$ and $\gamma^{\rm Coul}$ take form
\begin{align}
	\label{gamma_Coul_lin}
	&\gamma^{\rm Coul}_{\rm lin}={e^3 \tau^2 v_0^2/(\pi\hbar^2 \varepsilon_{\rm F}\omega) \over (1+\omega^2\tau^2)  [1+(\omega-\omega_{\rm pl}^2/\omega)^2\tau^2]}, \\
	&\gamma^{\rm SR}_{\rm lin}=-\gamma^{\rm Coul}_{\rm lin}{\Omega(2\Omega^4 + \Omega^2 +8)\over (\Omega^2+4)^2}.
\end{align}
Here, $\tau\propto\varepsilon_{\rm F}$ for Coulomb scattering and $\tau\propto1/\varepsilon_{\rm F}$ for short-range scattering.

The above equations show that the ratchet current is strongly enhanced by plasmonic effects in the frequency range close to the 2D plasmon. Owing to the lateral superlattice superimposed onto the tellurene, excitation of the plasmons becomes possible. According to Eq.~\eqref{chi_C_SR}, see  the factor in square brackets, the circular ratchet current has sharp peak at the plasmon frequency. 

\begin{figure}[t] 
	\centering
	\includegraphics[width=\linewidth]{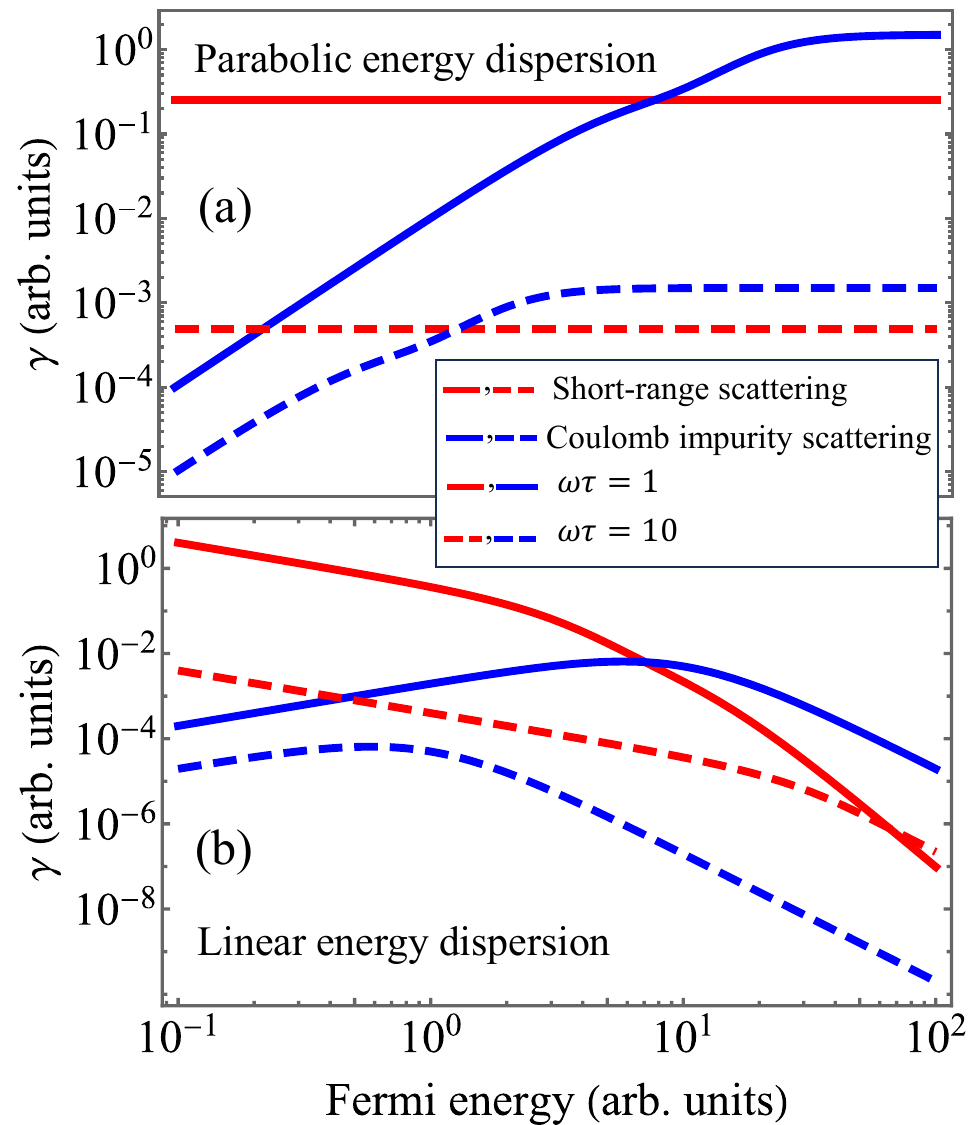}
	\caption{Fermi energy dependence of the circular ratchet current amplitude $\abs{\gamma}$ describing the behavior when the Fermi level lies deep in the conduction or valence band. Parabolic and linear energy dispersions and two elastic scattering mechanisms are considered. Solid and dashed curves are for $\omega\tau=1$ and 10, respectively. 
	}
	\label{fig_Fermi_en_depend}
\end{figure}

In the  structures under study, however, the 2D plasmon is far from the frequencies used in our experiments, therefore the ratchet current is described by Eqs.~\eqref{chi_C_SR} and~\eqref{chi_C_Coul} with $\omega_{\rm pl} \ll \omega$.
Under this assumption, the dependencies of $\gamma$ on the Fermi energy are shown in Fig.~\ref{fig_Fermi_en_depend} for two energy dispersions and short-range and Coulomb disorder potentials. It is seen from  Fig.~\ref{fig_Fermi_en_depend} that the circular ratchet current is strongly sensitive to both the type of energy dispersion and details of disorder scattering.

The experiments presented in this work are performed at room temperature. At a sufficiently low  $U_{\rm G}$ when the resistance has a maximum, the Fermi level lies in the band gap. Consequently, the free carriers are described by the Boltzmann statistics, and the circular ratchet current~\eqref{j_gamma} is determined by
\begin{equation}
	\label{chi_C_SR_Boltzmann}
	\gamma^{\rm SR} (T)= \gamma^{\rm SR} {N \pi \hbar^2\over m k_{\rm B}T},
\end{equation}
where $\gamma^{\rm SR}$ is the value~\eqref{chi_C_SR}, $N$ is the charge density, $T$ is the temperature, and $k_{\rm B}$ is the Boltzmann constant. For the linear energy dispersion this relation (between $\gamma^{\rm SR}_{\rm lin} (T)$ and  $\gamma^{\rm SR}_{\rm lin}$) also holds.

The gate voltage dependence of the circular ratchet amplitude presented in Fig.~\ref{fig4} follows from the theory presented above. At a sufficiently low  $U_{\rm G}$ the free carriers have Boltzmann statistics, and it follows from Eqs.~\eqref{chi_C_SR} and~\eqref{chi_C_SR_Boltzmann} that $\gamma \propto N$, therefore, the circular ratchet current monotonously increases with  $\abs{U_{\rm G}}$. This is true for both positive and negative  $U_{\rm G}$ provided the Fermi level is in the band gap. At high  $\abs{U_{\rm G}}$, by contrast, the carriers are degenerate. 
Note that in this case both electrons and holes have almost parabolic energy dispersion. Then, according to Fig.~\ref{fig_Fermi_en_depend}(a), $\gamma$  is a constant for short-range disorder or saturates for Coulomb scattering with increase of $\varepsilon_{\rm F}$. In both cases the linear increase of the circular ratchet current amplitude changes to a weaker dependence on  $U_{\rm G}$. Exactly this behavior takes place at high positive  $U_{\rm G}$ for both frequencies used in the experiments, see Fig.~\ref{fig4}.

The theory also explains the observed change of the circular ratchet current sign upon variation of the gate voltage clearly seen in Fig.~\ref{fig4}. Indeed, Eqs.~\eqref{chi_C_SR} and~\eqref{gamma_Coul_lin} reveal that the current is proportional to the third power of the carrier charge, which changes in the vicinity of the CNP. 
The slight shift of the zero point of the current from the CNP to negative gate voltages we attribute to the fact that 
electrons still substantially contribute to the current.
This is because  both electrons and holes are present at small $U_{\rm G}$, and electron's photocurrent prevails the hole's one.

And finally we discuss the ratio $r=J_{\rm circ}(f_1)/J_{\rm circ}(f_2)$ of the circular ratchet current amplitudes for two frequencies, $f_1=1.07$ and $f_2=2.02$~THz, 
see  Fig.~\ref{fig4}. 
The dependence of $r$ on the effective gate voltage is presented in Fig.~S6 of SI. For sufficiently large positive $U_{\rm G}$ (electrons), the ratio $r$ is slightly higher than 2, while for large negative $U_{\rm G}$ (holes) $r \approx 5.5$. 
It follows from Eqs.~\eqref{j_gamma},~\eqref{chi_C_SR} that for $\omega_{1,2}=2\pi f_{1,2} \gg \omega_{\rm pl}$, this ratio reads
\begin{equation}
r = {\omega_1(1+\omega_1^2\tau^2) \over \omega_2(1+\omega_2^2\tau^2)}.
\end{equation}
Using the relaxation times obtained from the mobilities (see SI for details), we get  $r \approx 2$ for electrons, which is close to the experimental value, and $r \approx 3.2$ for holes, which is somewhat smaller than the experimental one.
The latter could be caused by nonparabolicity of energy dispersion of the valence band which might result in a larger values of $\tau$ than those used in our estimations.
This difference requires further study.

\section{Summary and Outlook}


To summarize, we observed that the lateral asymmetric grating fabricated on top of  the 2D tellurene induces a circular ratchet current in the chiral axis direction $c$.    Due to the near-field diffraction, the grating periodically modulates the electric field of the incident radiation. Furthermore, the grating produces  the periodic electrostatic potential.  The developed semiclassical theory demonstrates that the combined action of these fields  on 2D carriers results in the DC helicity-driven electric current.   The DC current is controlled by the back gate voltage. The obtained Fermi energy dependence of the circular ratchet current describes well the experimental data for both positive and negative gate voltages.  In the former case, the current is carried by the chiral Weyl fermions, while in the latter case it is formed in the valence band with almost parabolic dispersion.  The current is detected at technologically important room temperature which paves the way for development of novel electronic devices. As a future task, since the circular ratchet current is controlled by the lateral asymmetry parameter, we expect that its magnitude can be substantially magnified by finding a special design that increases the value of the lateral parameter $\bm \Xi$. Moreover, tuning the structure parameters and radiation frequency to plasmonic resonance should result in the resonant circular current. Furthermore, decreasing the structure period and temperature would allow for a superlattice period smaller than the mean free path. This results in the formation of minibands, and in the new mechanisms of the circular ratchet current formation in the quantum-mechanical regime.

\acknowledgments
%
The financial support of the Deutsche Forschungsgemeinschaft (DFG, German Research Foundation) via Project-IDs 564981228 (MO 5219/1-1) and 521083032 (GA 501/19) is gratefully acknowledged.  Work of L.E.G. was funded by the German Research Foundation (DFG) as part of the German Excellence Strategy -- EXC3112/1 -- 533767171 (Center for Chiral Electronics). S.D.G. is grateful for the support of by the European Union through the ERC-ADVANCED grant TERAPLASM No. 101053716. 
Views and opinions expressed are, however, those of the author(s) only and do not necessarily reflect those of the European Union or the European Research Council Executive Agency. Neither the European Union nor the granting authority can be held responsible for them.

\bibliography{all_lib1.bib}

\newpage

\onecolumngrid

\newpage

\begin{center}
\makeatletter
{\large\bf{Supplementary Information for\\``\@title''}}
\makeatother
\end{center}

\let\oldsec\section

\renewcommand{\thesection}{S\arabic{section}}
\renewcommand{\section}[1]{\oldsec{#1}}
\renewcommand{\thepage}{S\arabic{page}}
\renewcommand{\theequation}{S\arabic{equation}}
\renewcommand{\thefigure}{S\arabic{figure}}

\setcounter{page}{1}
\setcounter{section}{0}
\setcounter{equation}{0}
\setcounter{figure}{0}

Supporting information includes additional data and details of the developed theory.

\twocolumngrid

\section{Additional data}
\label{appendixA1}

Additional data include:
\begin{itemize}
	
	\item Figure~S1 shows dependencies of the photocurrent excited by terahertz radiation in tellurene flakes with (upper panel) and without (lower panel) metal grating. The upper panel reveals a substantial circular photocurrent, manifesting itself by substantial difference between photoresponses excited with $\sigma^+$ and $\sigma^-$ radiation. In the grating free tellurene flake (lower panel), in contrast, the amplitudes of the photocurrent in response to the right- and left-handed circularly polarized radiation are almost equal.  The data are well fitted by Eqs.~5 and~6 of the main text (solid lines). For the data in the lower panel, the term proportional to the degree of circular polarization is settled to zero ($P_{\rm circ}=0$).  
	
	\item Figure S2 illustrates a power dependence of the photocurrent $J$ obtained in telurene flake with grating. It demonstrates a clear linear behavior upon the applied radiation intensity, $I$ being proportional to  square of the  radiation electric field $E^2$.

	\item Figure~S3 shows dependencies of the photocurrent excited by radiation with $f=2.02$~THz  on the phase angle $\varphi$. The data obtained for different effective gate voltages are presented as a waterfall plot. The data are well fitted by Eqs.~5 and~6 of the main text (solid lines).

	\item Figures~S4 and S5 present the effective gate voltage dependencies of the  coefficients used for fitting of phase angle dependencies in Figs.~2  and~5 of the main text and Fig.~S3 of the SI.
	
		\item Figure~S6 shows the ratio $r$ of the circular ratchet current amplitudes for  $f=1.07$ and $f=2.02$~THz as a function of $U_{\rm G}$. For the theoretical estimates we used for electrons $\mu_n=200$~cm$^2$/Vs (experiment), $m_n=0.1~m_0$~\cite{Qiu2020}, which yields $\tau=1.3 \times 10^{-14}$~s. For holes $\mu_p=520$~cm$^2$/Vs (experiment) and the reduced mass $m_p=0.19~m_0$~\cite{Couder1969}, which yields $\tau=5.6 \times 10^{-14}$~s.

\end{itemize}

\begin{figure}[h] 
	\centering
	\includegraphics[width=\linewidth]{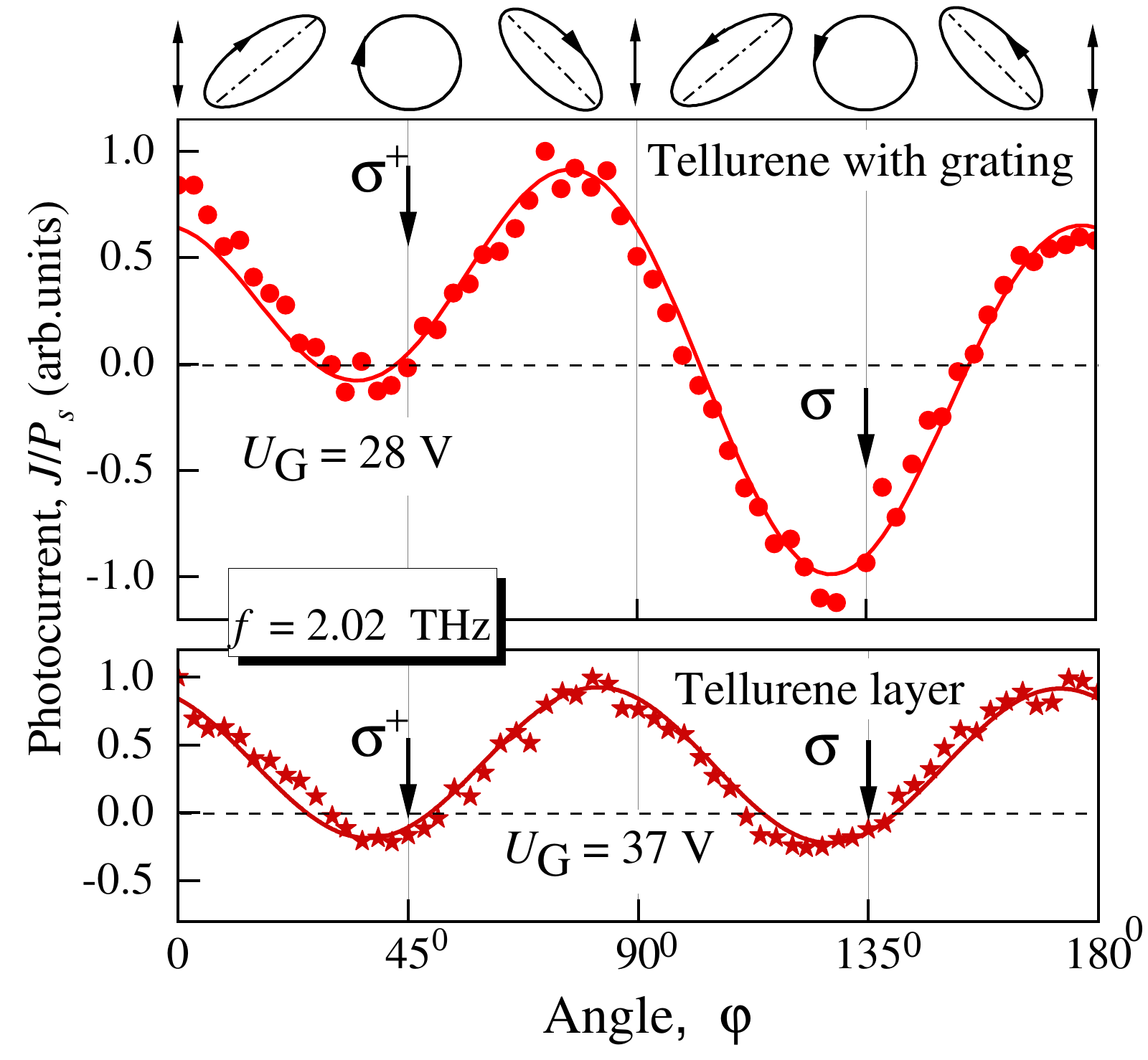}
	\caption{The normalized photocurrent, $J/P_{\rm s}$, as a function of the lambda-quarter rotation angle, $\varphi$. The data are shown for a  radiation frequency $f=2.02$~THz. The solid lines represent the corresponding fits according to Eqs.~5 and 6 of the main text. Vertical arrows indicate the right- ($\sigma^+$) and left-handed ($\sigma^-$) circular polarization states and the ellipses on top illustrate the polarization states at various angles~$\varphi$. 
	}
	\label{figS1}
\end{figure}


\begin{figure}[h] 
	\centering
	\includegraphics[width=\linewidth]{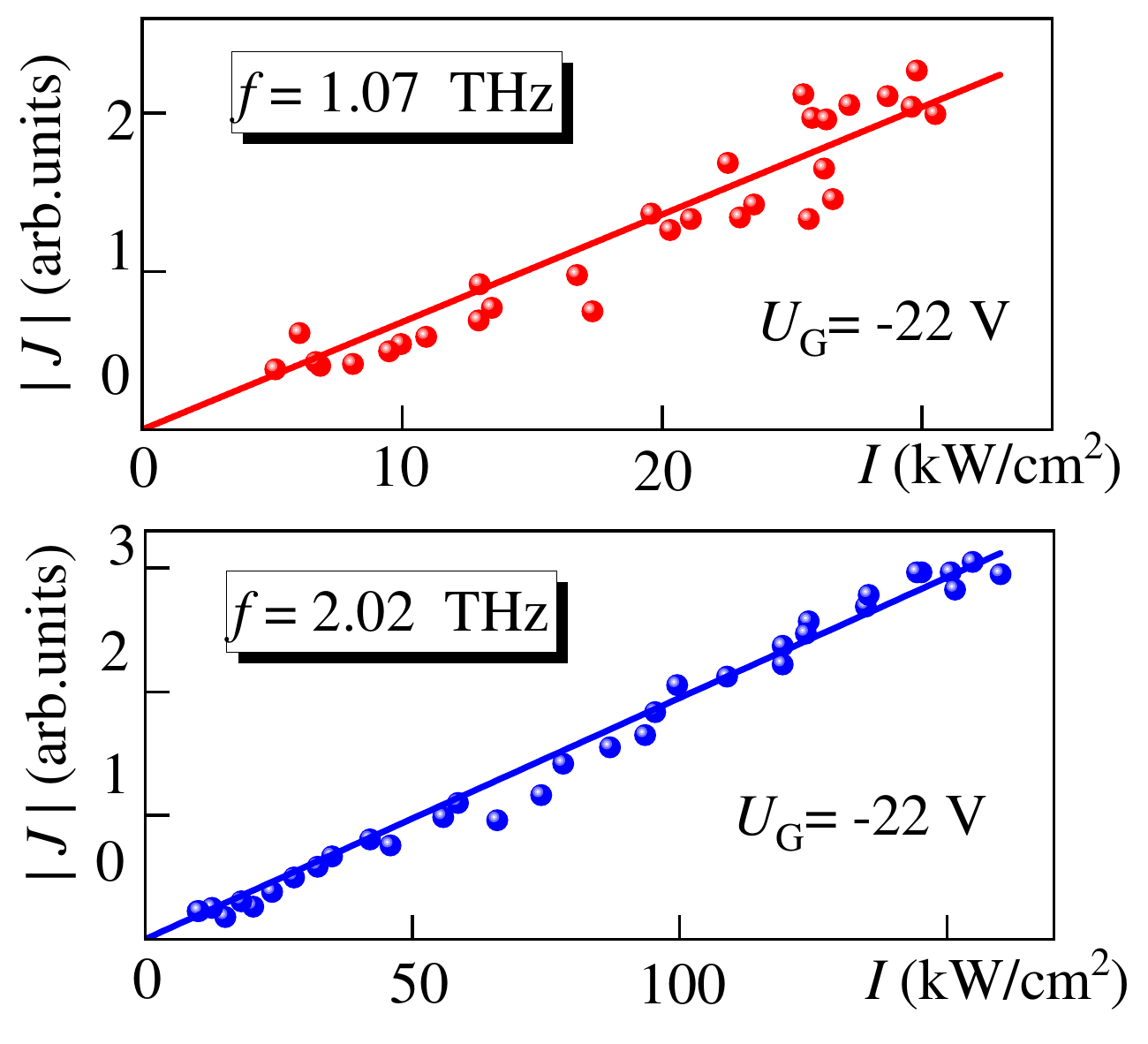}
	\caption{Exemplary dependencies of the  photocurrent $J$   on  the radiation intensity $I = P / S_{\rm spot} \propto E^2$. The data are obtained for  $f=1.07$ (upper panel) and $f=2.02$~THz (lower panel). 
	}
	\label{figS2}
\end{figure}

\begin{figure}[h] 
	\centering
	\includegraphics[width=\linewidth]{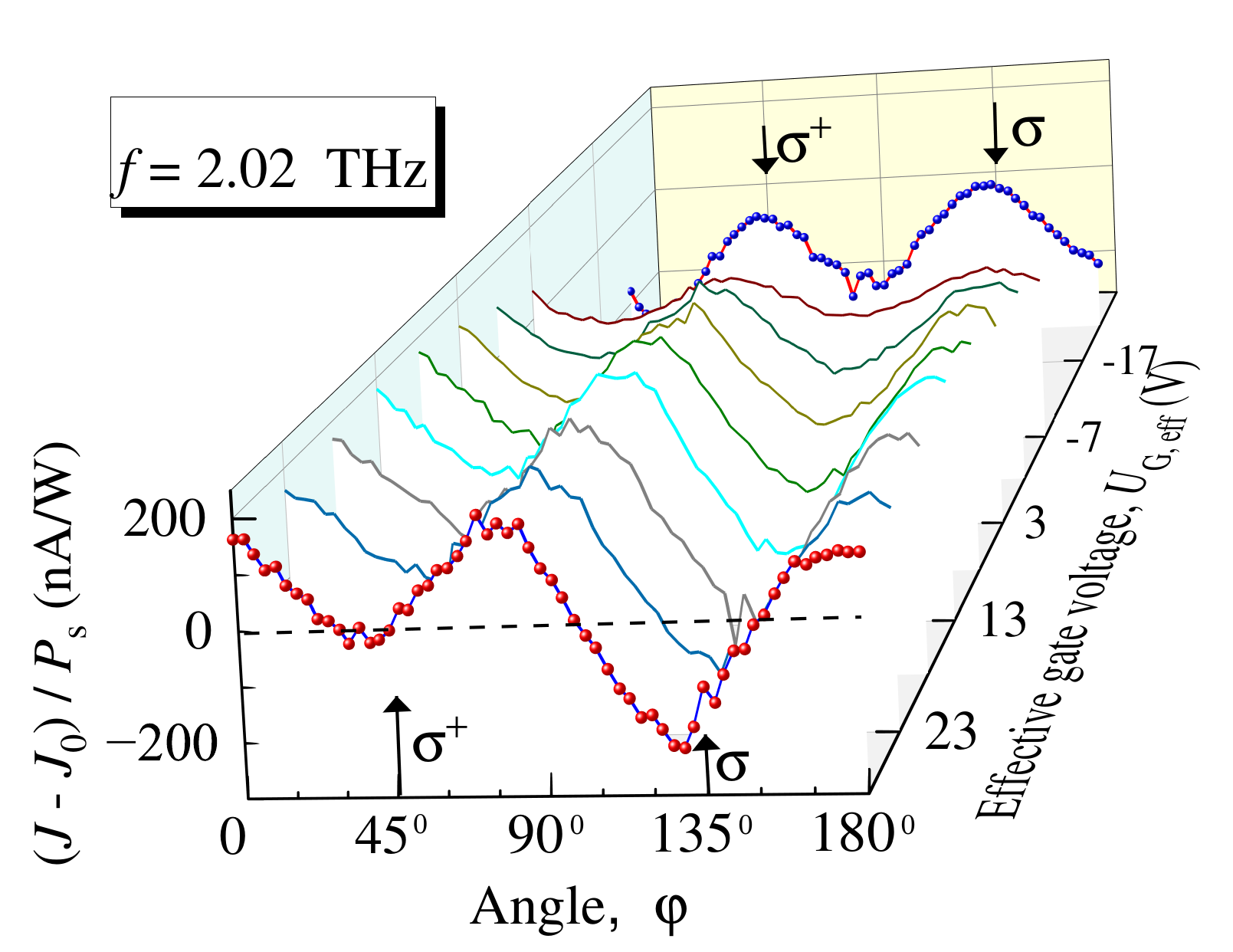}
	\caption{Dependencies of the photocurrent excited by radiation with $f=2.02$~THz on the rotation angle $\varphi$ of the $\lambda/4$ wave plate. The waterfall plot shows the data for different effective back gate voltages ranging from  $U_{\rm G} =-20$ to  $U_{\rm G} = 26$~V.  The photoresponse is normalized to the laser power $P_{\rm s}$  incident on the sample.  The vertical arrows  mark the right-handed ($\sigma^+$) and left-handed ($\sigma^-$) circular polarization states, respectively. The solid lines represent the corresponding fits according to Eqs.~5 and 6 of the main text. The fit coefficients are plotted in Fig.~S5.
	}
	\label{figS3}
\end{figure}

\begin{figure}[h] 
	\centering
	\includegraphics[width=\linewidth]{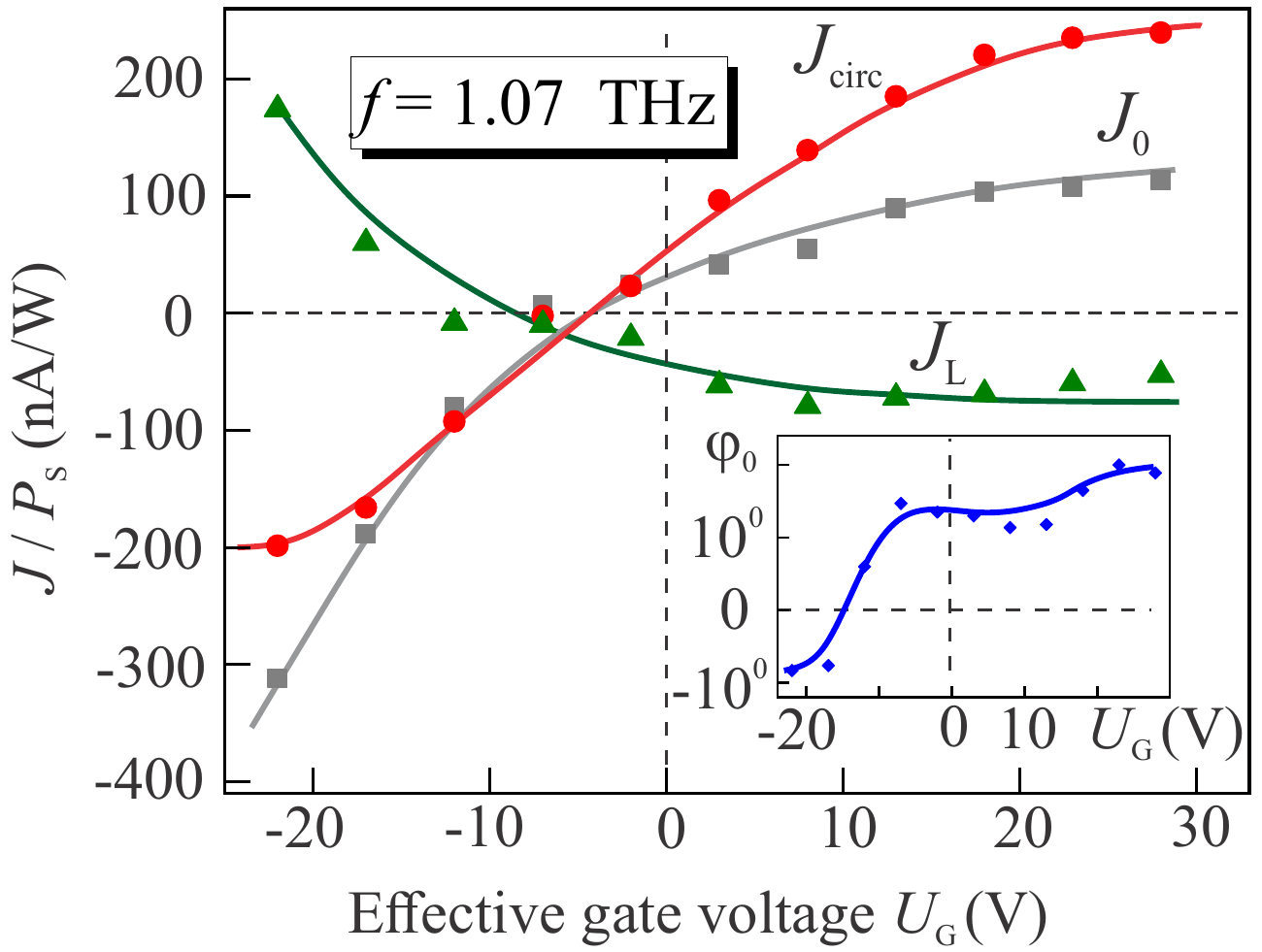}
	\caption{Effective gate voltage dependencies of the  parameters used for fitting of the angle $\varphi$ dependencies in Figs. 2  and 5 of the main text: Three photocurrent contributions in response by radiation with $f=1.07$~THz are circular $J_{\rm circ}$, polarization independent $J_0$, and linear $J_{\rm L}$, see Eq.~(6). The inset shows the gate dependence of the phase angle $\varphi_0$  of the linear ratchet current. 
	}
	\label{figS4}
\end{figure}

\begin{figure}[h] 
	\centering
	\includegraphics[width=\linewidth]{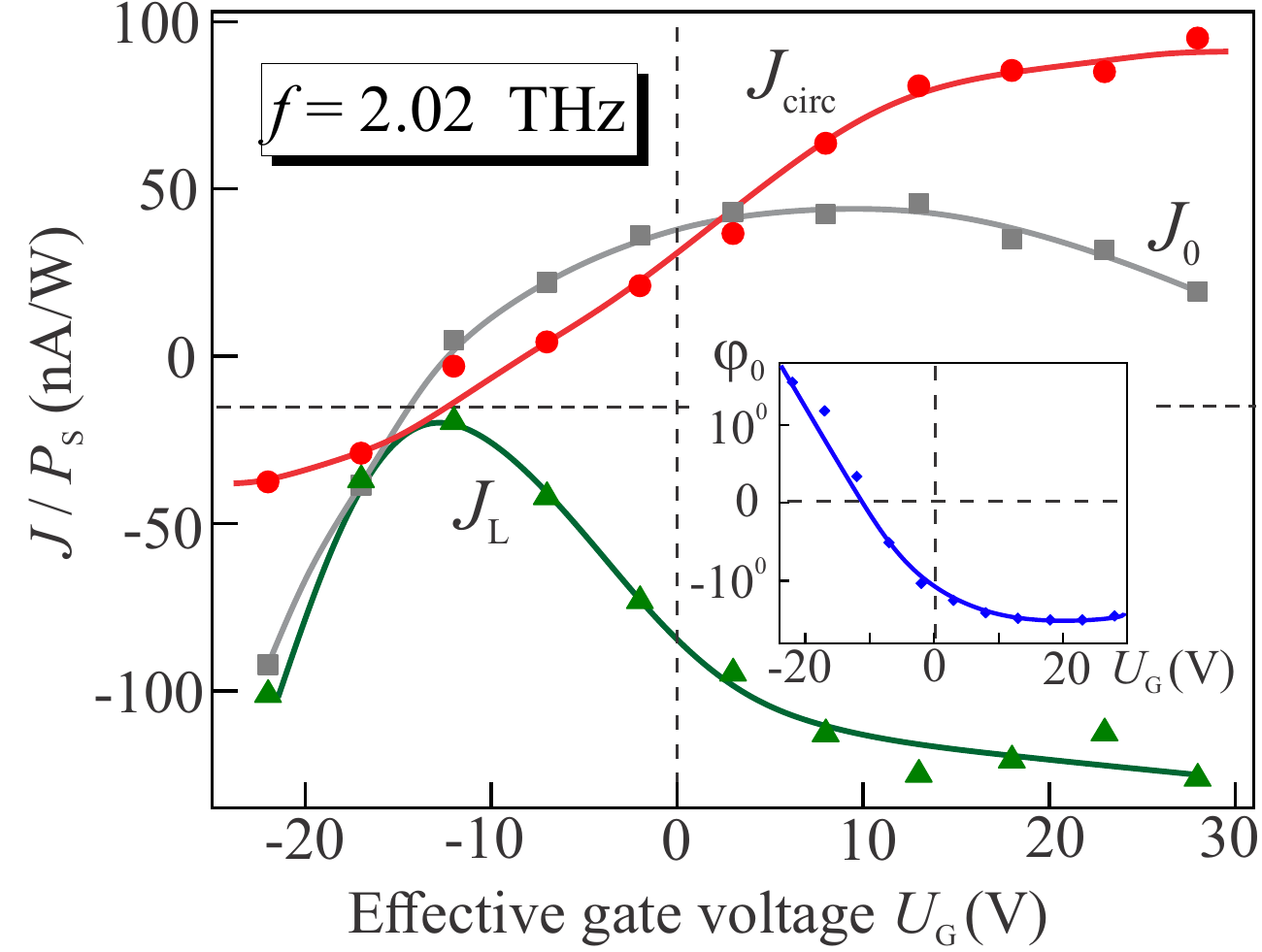}
	\caption{Effective gate voltage dependencies of the  parameters used for fitting of  the angle $\varphi$ dependencies in Fig.~S3 of SI: Three photocurrent contributions in response by radiation with $f=2.02$~THz are circular $J_{\rm circ}$, polarization independent $J_0$, and linear $J_{\rm L}$, see Eq.~(6). The inset shows the gate dependence of the phase angle $\varphi_0$  of the linear ratchet current. 
	}
	\label{figS5}
\end{figure}

\begin{figure}[h] 
	\centering
	\includegraphics[width=\linewidth]{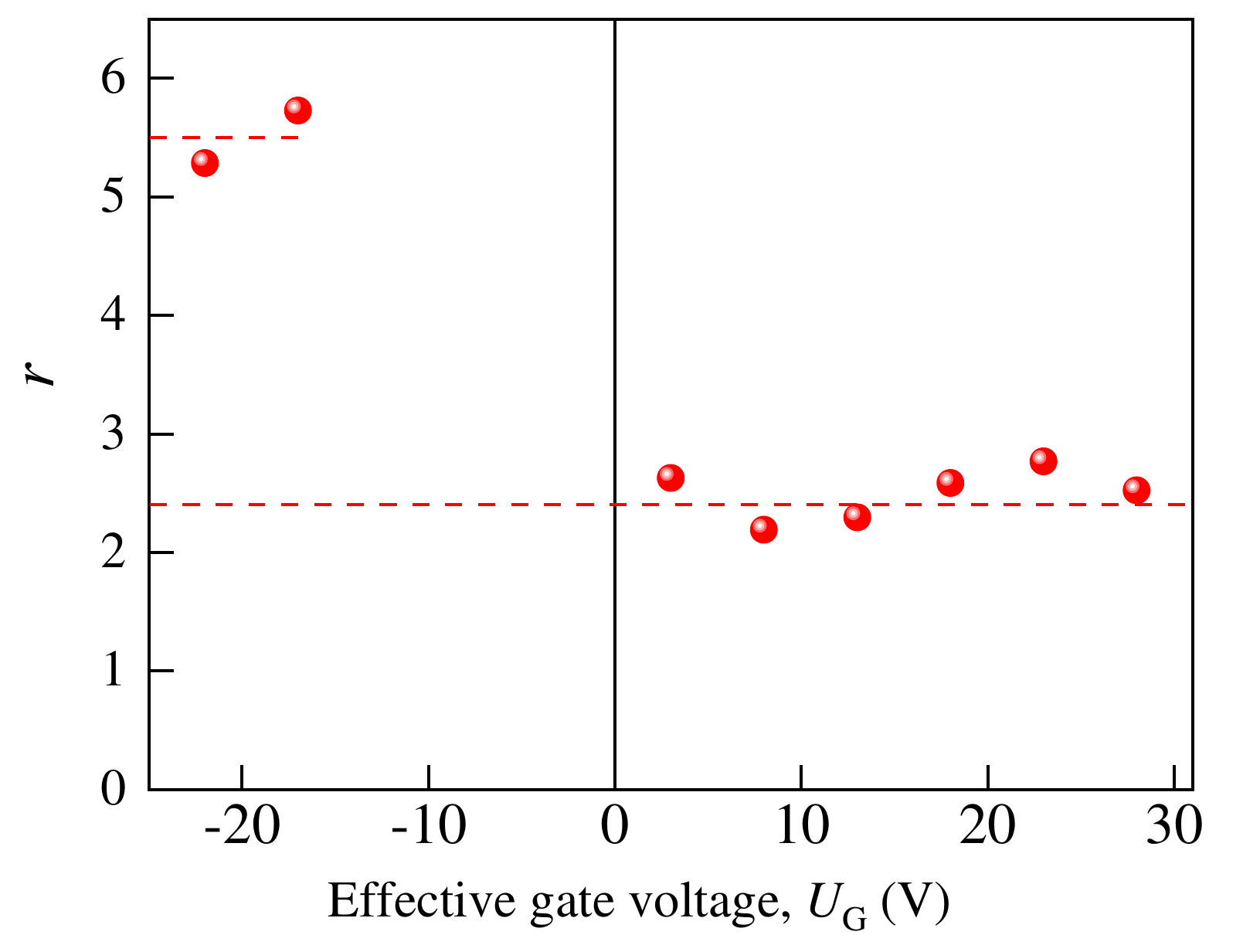}
	\caption{The ratio of the circular ratchet current amplitudes $r=J_{\rm circ}(f_1)/J_{\rm circ}(f_2)$ for  $f_1=1.07$ and $f_2=2.02$~THz as a function of $U_{\rm G}$. The amplitudes $J_{\rm circ}(f_1)$ and $J_{\rm circ}(f_2)$ are taken from Fig.~4 of the main text.
	}
	\label{figS6}
\end{figure}

\section{Theory}

The ratchet current is proportional to the components $\Xi_{x,y}$ of  the structure asymmetry parameter
\begin{equation}
	\bm \Xi = \overline{E^2(x,y)\bm \nabla V(\bm r)}.
\end{equation}
Despite both the static force and the radiation intensity modulation have zero mean values, their product $\bm \Xi$ is nonzero in the studied structure of the $C_1$ point symmetry. The result of calculations reads~\cite{Yahniuk2024}
\begin{equation}
	j_x= \Xi_y \gamma P_{\rm circ} +\Xi_x \left( \chi_L  {P_L} + \chi_0 \right) +\Xi_y \tilde{\chi}_L {\tilde{P}_L}.
\end{equation}
Here $P_{\rm circ}$ is the radiation helicity, and $P_L$, $\tilde{P}_L$ are the Stokes parameters defining the degrees of the linear polarization,  and the factors $\gamma$, $\chi_L$, $\tilde{\chi}_L$, $\chi_0$ describe the circular, linear and polarization-independent  ratchet effects, respectively. These factors are independent of the periodic potential and radiation intensity, being determined by the properties of 2D carriers and radiation frequency~\cite{Nalitov2012,Moench2023b}.

For the helicity-driven ratchet current $j_x^{\rm circ}= \Xi_y \gamma P_{\rm circ}$ we have
\begin{equation}
	\label{SI_gamma}
	\gamma = \int_0^\infty \dd \varepsilon [-f_0'(\varepsilon)] \gamma_1(\varepsilon),
\end{equation}
where $f_0(\varepsilon)$ is the Fermi-Dirac  distribution, and prime here and below denotes differentiation over the energy $\varepsilon$. The function $\gamma_1(\varepsilon)$ is given by~\cite{Nalitov2012}
\begin{equation}
	\label{SI_gamma1}
	\gamma_1(\varepsilon) ={e^3 v^2\over 2\pi\hbar^2} \qty({\rm Re}S_2 - {\rm Im}S_1),
\end{equation}
where the velocity $\bm v=\partial \varepsilon_{\bm p}/\partial \bm p$.

For the parabolic energy dispersion $\varepsilon_{\bm p}=p^2/(2m)$, $v^2=2 \varepsilon/m$, we have~\cite{Nalitov2012}
\begin{align}
	S_{1} &= \varepsilon \tau_1' \tau_{2\omega} \tau_{1\omega}' - {1 \over 2\varepsilon} \left( \varepsilon^2 \tau_1' \tau_{2\omega} \tau_{1\omega}  \right)', \nonumber \\
	S_{2} &=	{\left( \tau_1 \varepsilon \right)'  \left( \tau_{1 \omega} \varepsilon \right)' \over \omega \varepsilon} - {\left[ \left( \tau_1 \varepsilon \right)' \tau_{1\omega} \varepsilon \right]' \over 2\omega \varepsilon} .  \label{SI_S12_parab}
\end{align}
Here the energy-dependent relaxation times are defined via the elastic scattering probability $W_{\bm p' \bm p}$ as follows
\begin{equation}
	{1\over \tau_{n}(\varepsilon_p)} = \sum_{\bm p'}W_{\bm p' \bm p}(1-\cos{n\theta}), \quad n=1,2,
\end{equation}
where $\theta$ is the angle between $\bm p$ and $\bm p'$, and
the complex relaxation times are
\begin{equation}
	\tau_{n\omega} = {\tau_n\over 1-i\omega\tau_n}.
\end{equation}
Note that, at degenerate statistics, the transport relaxation time $\tau$ equals to $\tau_1$ at $\varepsilon=\varepsilon_{\rm F}$, and  $\gamma=\gamma_1(\varepsilon_{\rm F})$.

For short-range (SR) disorder, the scattering times  $\tau_1=\tau_2$ are independent of energy and equal to $\tau$. Then Eqs.~\eqref{SI_gamma1},~\eqref{SI_S12_parab} yield a constant $\gamma_1$ given by
\begin{equation}
	\label{SI_chi_C_SR}
	\gamma_1^{\rm SR} 
	= {e^3 \tau^2\over 2\pi\hbar^2 m\omega [1+(\omega\tau)^2]}.
\end{equation}

For long-range Coulomb (Coul) disorder scattering we have $\tau_1=\tau{\varepsilon\over \varepsilon_{\rm F}}$, and $\tau_2=\tau_1/2$. Therefore we obtain
\begin{equation}
	\label{SI_chi_C_Coul}
	\gamma^{\rm Coul}(\varepsilon) = \gamma^{\rm SR}{3\Omega^6 + 29\Omega^4 + 4\Omega^2 +32\over (1+\Omega^2)(\Omega^2+4)^2},
\end{equation}
where $\Omega=\omega\tau \propto \varepsilon$.

Inclusion of the plasmon resonance is performed as follows
\begin{equation}
	\gamma \to \gamma {1+(\omega\tau)^2\over1+(\omega-\omega_{\rm pl}^2/\omega)^2\tau^2}.
\end{equation}

For the linear  energy dispersion $\varepsilon_{\bm p}=v_0p$, $v=v_0$, the coefficients $S_{1,2}$ in Eq.~\eqref{SI_gamma1} are defined by the following expressions~\cite{Nalitov2012}
\begin{align} \label{SI_S}
	S_{1,{\rm lin}} &= \varepsilon^3 \left( \tau_1 \over \varepsilon \right)'  \tau_{2\omega} \left( \tau_{1\omega} \over \varepsilon \right)' - {\left[ \varepsilon^2 \left( \tau_1 \over \varepsilon \right)' \tau_{2\omega} \tau_{1\omega}  \right]' \over 2} , 
	\nonumber \\
	S_{2,{\rm lin}} &=	{\left( \tau_1 \varepsilon \right)'  \left( \tau_{1 \omega} \varepsilon \right)' \over \omega \varepsilon} - {\left[ \left( \tau_1 \varepsilon \right)' \tau_{1\omega} \right]'\over 2\omega } . 
\end{align}
For short-range disorder, the scattering times  $\tau_1^{\rm SR}=\tau{\varepsilon_{\rm F}\over \varepsilon}$, and $\tau_2^{\rm SR}=\tau_1^{\rm SR}/2$, while for  Coulomb scattering we have $\tau_1^{\rm Coul}=\tau{\varepsilon\over \varepsilon_{\rm F}}$, and $\tau_2^{\rm Coul}=3\tau_1^{\rm Coul}$.
This yields
\begin{align}
	\label{SI_gamma_Coul_lin}
	&\gamma^{\rm Coul}_{\rm lin}={e^3 \tau^2 v_0^2/(\pi\hbar^2 \varepsilon_{\rm F}\omega) \over (1+\omega^2\tau^2)  [1+(\omega-\omega_{\rm pl}^2/\omega)^2\tau^2]}, \\
	&\gamma^{\rm SR}_{\rm lin}=-\gamma^{\rm Coul}_{\rm lin}{\Omega(2\Omega^4 + \Omega^2 +8)\over (\Omega^2+4)^2}.
\end{align}

For non-degenerate statistics, the energy averaging~\eqref{SI_gamma} should be performed with the Boltzmann distribution. At short-range scattering, $\gamma^{\rm SR} $ is a constant, and we get
\begin{equation}
	\label{SI_chi_C_SR_Boltzmann}
	\gamma^{\rm SR} (T)= \gamma^{\rm SR} {N \pi \hbar^2\over m k_{\rm B}T},
\end{equation}
where 
$N$ is the charge density, 
$T$ is the temperature, and $k_{\rm B}$ is the Boltzmann constant.
For the linear energy dispersion this relation also holds.

The dependencies of $\gamma$ on the Fermi energy is shown in Fig.~7 of the main text for two energy dispersions and short-range and Coulomb disorder potentials.
\end{document}